# Two-Way Latent Grouping Model for User Preference Prediction

**Eerika Savia** and **Kai Puolamäki**
[1]Laboratory of Computer and
Information Science
Neural Networks Research Centre
Helsinki University of Technology
P.O. Box 5400
FI-02150 TKK, Finland

**Janne Sinkkonen**
Xtract Ltd.
Helsinki, Finland

**Samuel Kaski**[1,2]
[2]Department of Computer Science
P.O. Box 68
FI-00014 University of Helsinki
Finland

## Abstract

We introduce a novel latent grouping model for predicting the relevance of a new document to a user. The model assumes a latent group structure for both users and documents. We compared the model against a state-of-the-art method, the User Rating Profile model, where only users have a latent group structure. We estimate both models by Gibbs sampling. The new method predicts relevance more accurately for new documents that have few known ratings. The reason is that generalization over documents then becomes necessary and hence the two-way grouping is profitable.

## 1 INTRODUCTION

This work has been done as a part of a *proactive information retrieval* project (PRIMA, 2003–2005) that aims at estimating relevance of new documents to users based on both explicit and implicit user feedback. A proactive system adapts to the interests of the user, inferred from implicit feedback. User feedback by explicitly pointing out relevant documents might sometimes be more accurate, but users often consider giving such feedback too laborious. The usability and accuracy of information retrieval applications would therefore be greatly enhanced by complementing explicit feedback with more readily available implicit feedback signals measured from the user interface.

As implicit feedback typically is noisier than explicit ratings, efficient generalization over users and documents becomes necessary. In this paper we introduce a two-way latent grouping model that predicts binary relevances for user-document pairs.

Traditionally, user preferences have been predicted using so-called collaborative filtering methods, where the predictions are based on the opinions of similar-minded users. Collaborative filtering is needed when the task is to make personalized predictions but there is not yet sufficient amount of data about the user's personal interests. Then the only possibility is to generalize over users, for instance by grouping them into like-minded user groups. The early methods were memory-based; predictions were made by identifying a set of similar users, and using their preferences fetched from memory. See, for instance (Shardanand and Maes, 1995) and (Konstan et al., 1997). Model-based approaches are justified by the exponentially increasing time and memory requirements of the memory-based techniques. Recent work includes probabilistic and information-theoretic models, for instance (Hofmann, 2004; Wettig et al., 2003; Zitnick and Kanade, 2004; Jin and Si, 2004). An interesting family of models are the latent component models, which have been successfully used in collaborative filtering (Pritchard et al., 2000; Hofmann, 2004; Blei et al., 2003; Marlin, 2004). In these models, each user is assumed to belong to one or many latent user groups that explain her preferences.

As a collaborative filtering system has to rely on the past experiences of the users, it will have problems when assessing new documents not seen yet by most of the users. To tackle this problem we propose a model that generalizes over documents as well. We go one step further from the state-of-the-art probabilistic models which have a latent structure for the users, and introduce a similar latent structure for the documents as well.

We will test the model for two different kinds of data sets. The model is generally applicable, but since our main application area is proactive information retrieval, we will refer to the items as documents. In both of the tests the performance of our model is compared to a state-of-the-art method, the User Rating Profile model, URP (Marlin, 2004), which we also evaluated by Gibbs sampling. In the work of Marlin, the URP model, which was evaluated with variational methods, outperformed the other latent topic models. Our first study involves a set of about 500,000 votes given by 679 members of British parliament on roughly 1,200 issues. In the second study, the data were gathered in a controlled experiment where the test subjects browsed through a set of titles of scientific articles, and chose the most interesting ones via a web form. The data consisted of 25 users' opinions on 480 articles.

This paper is structured as follows. We first introduce our two-way grouping model and the models used in the experimental as comparisons, including the baseline models. After that we discuss the differences and similarities of our model with the related models. We then proceed to describe the experimental setup and the results.

## 2 METHODS

### 2.1 TWO-WAY GROUPING MODEL

We introduce a model that clusters users into user groups and documents into document clusters, in order to generalize relevance over both groupings. Each user may have several "attitudes," that is, belong to different groups during different relevance evaluations, and likewise each document may have several "aspects." These are modeled as probabilistic soft assignments. Our notation is summarized in Table 1.

Table 1: Notation

| SYMBOL | DESCRIPTION |
|---|---|
| $u$ | user index |
| $d$ | document index |
| $r$ | binary relevance |
| $u^*$ | user group index |
| $d^*$ | document cluster index |
| $Z$ | user attitude in URP |
| $N_U$ | number of users |
| $N_D$ | number of documents |
| $N$ | number of triplets $(u, d, r)$ |
| $K_U$ | number of user groups |
| $K_D$ | number of document clusters |
| $\mathcal{D}$ | observed data |

The model generates rating triplets (user, document, rating), or $(u, d, r)$, with binary relevances $r$ as follows (see Figure 1):

1A) For the whole user collection, a vector of Multinomial parameters $\theta_U$ is drawn from Dirichlet($\boldsymbol{\alpha}_{u^*}$). The parameter vector $\theta_U$ contains the probability for each user group $u^*$ to occur.

2A) For each user group $u^*$, a vector of Multinomial parameters $\boldsymbol{\beta}_U(u^*)$ is drawn from Dirichlet($\boldsymbol{\alpha}_U$). The parameter vector $\boldsymbol{\beta}_U(u^*)$ contains the probability for each user to belong to user group $u^*$.

3A) A user group $u^*$ is drawn from Multinomial($\theta_U$). As the user group is fixed the corresponding parameters $\boldsymbol{\beta}_U(u^*)$ can be selected and a user $u$ is drawn from Multinomial($\boldsymbol{\beta}_U(u^*)$).

1B) Symmetrically, for the whole document collection, a vector of Multinomial parameters $\theta_D$ is drawn from Dirichlet($\boldsymbol{\alpha}_{d^*}$). The parameter vector $\theta_D$ contains the probability for each document cluster $d^*$ to occur.

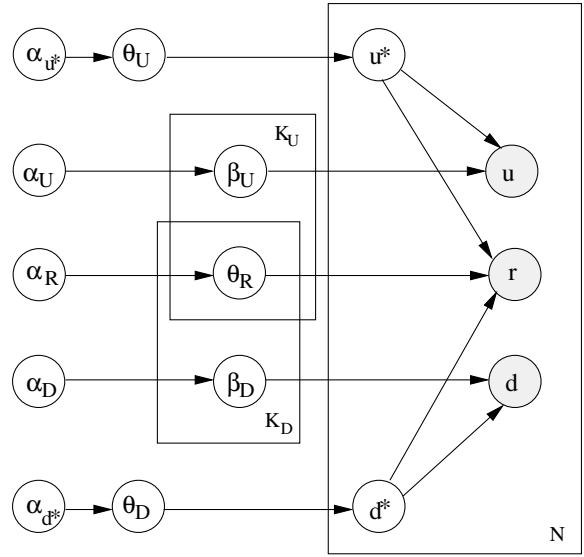

Figure 1: A Graphical Model Representation of Our Two-Way Grouping Model. The grey circles indicate observed values. The boxes are "plates" representing replicates; the value in the corner of each plate is the number of replicates. The rightmost plate represents the repeated choice of $N$ (user, document, rating) triplets. The plate labeled with $K_U$ represents the multinomial models of the $K_U$ different user groups. The plate labeled with $K_D$ represents the multinomial models of the $K_D$ different document clusters.

2B) For each document cluster $d^*$, a vector of Multinomial parameters $\boldsymbol{\beta}_D(d^*)$ is drawn from Dirichlet($\boldsymbol{\alpha}_D$). The parameter vector $\boldsymbol{\beta}_D(d^*)$ contains the probability for each document to belong to the document cluster $d^*$.

3B) A document cluster $d^*$ is drawn from Multinomial($\theta_D$). As the document cluster is fixed the corresponding parameters $\boldsymbol{\beta}_D(d^*)$ can be selected and the document $d$ is drawn from Multinomial($\boldsymbol{\beta}_D(d^*)$).

4) For each cluster pair $(u^*, d^*)$, a vector of Binomial parameters $\theta_R(u^*, d^*)$ is drawn from Dirichlet($\boldsymbol{\alpha}_R$). The parameters $\theta_R(u^*, d^*)$ contain the probability of the user group $u^*$ to consider the document cluster $d^*$ relevant (or irrelevant).

5) For each cluster pair $(u^*, d^*)$, a binary relevance $r$ is drawn from Binomial($\theta_R(u^*, d^*)$).

The model parameters are simulated with Gibbs sampling. As a very brief tutorial, the model parameters are sampled one at a time, conditional on the current values of all the other parameters. One iteration step consists of sampling all the parameters once, in a fixed order. The observed data consists of triplets $(u, d, r)$. For each iteration step of sampling, $m$, we get a sample of all parameters of the model, denoted by $\boldsymbol{\psi}^{(m)}$. Asymptotically, the sampled parameters $\boldsymbol{\psi}^{(m)}$ satisfy $\boldsymbol{\psi}^{(m)} \sim P(\boldsymbol{\psi} \mid \mathcal{D})$.

Each sample of parameters generates a matrix of probabilities $P(r, u, d \mid \boldsymbol{\psi}^{(m)})$. The prediction of relevance, $P(r \mid u, d)$, can be computed from these by the Bayes rule. As the final prediction we use the mean of the predictions over the $M$ Gibbs iterations,

$$\begin{aligned} P(r \mid u, d) &= \mathbb{E}_{P(\boldsymbol{\psi}|\mathcal{D})}\left[P(r \mid u, d, \boldsymbol{\psi})\right] \\ &\approx \frac{1}{M}\sum_{m=1}^{M} P(r \mid u, d, \boldsymbol{\psi}^{(m)}). \end{aligned} \quad (1)$$

The sampling formulas are presented in the Appendix.

## 2.2 USER RATING PROFILE MODEL

URP is a generative model which generates a binary rating $r$ for a given (user, document) pair.[1] It was originally optimized with variational Bayesian methods (*variational URP*), including also maximum likelihood estimates. We replaced the optimization with the potentially more accurate Markov chain Monte Carlo sampling from the full posterior distribution (*Gibbs URP*). This is expected to improve estimates especially for the small numbers of known ratings in our application. We estimate the posterior predictive distribution $P(r|u, d, \mathcal{D})$ by Gibbs sampling where $\mathcal{D}$ denotes the training data consisting of observations $(u, d, r)$. The model assumes that there are a number of latent user groups whose preferences on the documents vary. The users belong to these groups probabilistically, into different groups at different times. Alternatively, the groups can be interpreted as different "attitudes" of the user, and the attitude may be different for different documents.

The generative process proceeds according to the following steps (see also Figure 2):

1) For each user, a vector of multinomial parameters $\boldsymbol{\theta}(u)$ is drawn from Dirichlet($\boldsymbol{\alpha}$). The parameter vector $\boldsymbol{\theta}(u)$ contains the probabilities for the user to have different attitudes $Z$, that is, to belong to different user groups $Z$.

2) For each user $u$, a user group or attitude $Z$ is drawn for each document $d$, from the user's Multinomial($\boldsymbol{\theta}(u)$). The value of $Z$ in effect selects the parameters $\boldsymbol{\beta}(Z, d)$ from the set of parameters in the node labeled by $\beta$ in Figure 2.

3) For each (user group, document) pair $(Z, d)$, a vector of binomial parameters $\beta(Z, d)$ is drawn from Dirichlet($\boldsymbol{\alpha}_\beta(Z, d)$). The parameters $\beta(Z, d)$ define the probability of the user group $Z$ to consider document $d$ relevant (or irrelevant).

4) For each pair $(Z, d)$, a binary relevance value $r$ is drawn from the Binomial($\beta(Z, d)$).

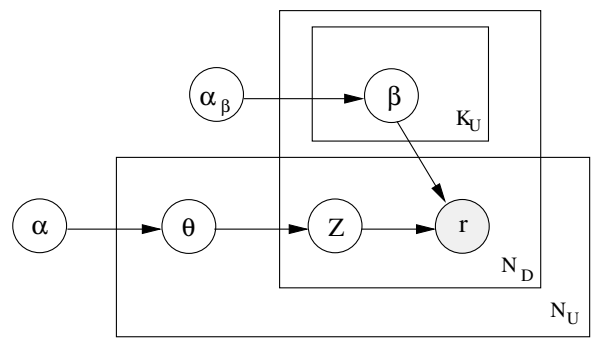

Figure 2: A Graphical Model Representation of URP. The grey circle indicates an observed value. The boxes are "plates" representing replicates and the value at the corner of each plate indicates the number of replicates. The lowest plate represents users and the plate labeled with $N_D$ represents the repeated choice of user group for each document. The plate labeled with $K_U$ represents the multinomial models of relevance for the $K_U$ different user groups.

---
[1] Note that the model allows also multiple-valued ratings if the binomial is replaced with a multinomial.

## 2.3 NAIVE MODEL AND DOCUMENT FREQUENCY MODEL

We implemented two simple models to give baseline results. The *naive model* always predicts the same relevance value for $r$, according to the more frequent value in the training set. The prediction of the naive model was $r = 0$ for the scientific articles and $r = 1$ for the parliament votings.

The *document frequency model* does not take into account differences between users or user groups. It simply models the probability of a document being relevant as the frequency of $r = 1$ in the training data for the document:

$$P(r = 1 \mid d) = \frac{\sum_u \#(u, d, r = 1)}{\sum_{u,r} \#(u, d, r)} \quad .$$

## 2.4 COMPARISON TO OTHER MODELS

The models most closely related to our two-way grouping model are the so-called latent topic models, especially Hofmann's probabilistic latent semantic analysis (pLSA) (Hofmann, 2004), Latent Dirichlet Allocation (LDA) (Blei et al., 2003) also known as multinomial PCA or mPCA (Buntine, 2002), and the already introduced URP (Marlin, 2004). The main differences to these three models are discussed in this Section.

The three models have subtle differences, but in all of them each user is assigned an individual multinomial distribution with parameters $\boldsymbol{\theta}$, and the latent user group $Z$ is sampled again for each document[2]. Each user can therefore belong to many groups with varying degrees. In our model both *users* and *documents* can belong to many latent groups, much in the same way as users in these three models.

In pLSA and URP each user group has a set of multinomials $\text{Mult}(\beta)$ for all documents—they immediately determine the probabilities of ratings once the user group has been generated. Each mPCA user group, on the contrary, has a multinomial $\text{Mult}(\beta)$ over the *documents*. Hence, mPCA could be interpreted as a model where the most probably occurring documents are the ones that have the largest probability of being relevant. Thus, multinomial PCA can be interpreted as a binary relevance model but it cannot represent multiple-valued ratings. URP can be seen as an extension to mPCA with one extra dimension in the parameter matrix $\boldsymbol{\beta}$ to represent the different rating values. In our model the relevance is assumed to depend only on the latent groups, that is, there is a probability distribution of different ratings, $\text{Mult}(\theta_R)$, for each (user group, document cluster) pair.

In addition to being two-way, our model differs from URP in that the users $u$ and documents $d$ are explicitly generated. In other words, the margins $P(u)$ and $P(d)$ are estimated from data. In contrast, URP contains no generative process for the users or documents.

Finally, although our model is not far from the diverse set of so-called biclustering models (Tanay et al., 2005; Madeira and Oliveira, 2004; Wettig et al., 2003), we aim at prediction instead of clustering, and therefore it is enough that the latent structure makes the predictions accurate. Because the model is evaluated by sampling, we do not even obtain a single explicit set of clusters.

## 3 DATA

### 3.1 PARLIAMENT DATA

The model was first empirically tested on a publicly available data set of votings of the British parliament during year 1997. The data set consisted of 514,983 votes given by the members of parliament on 1,272 issues. We predict the votes for previously unseen (user, issue) pairs, where the users are the members of the parliament (MP).

#### 3.1.1 Missing Data

In the parliament data we do not have a "yes" or "no" answer for all the (user, issue) pairs, i.e., about 40 % of all possible votes are missing. The fact that a member of parliament has not voted "yes" or "no" on a particular issue may either be due to her not being present at the voting, or her tactical reasons not to take a stand in the matter. In either case, the vote is not missing at random. From the modeling perspective we could assign a "rating," say -1, to all the missing votes and make the sparse data matrix full. This would, however, notably increase the computational load. Fortunately, some of this information is effectively taken into account by modeling the user margins $\text{P}(u \mid \mathcal{D})$ and the document margins $\text{P}(d \mid \mathcal{D})$.

### 3.2 SCIENTIFIC ARTICLES DATA

Our other data set was gathered in a controlled experiment where the test subjects browsed through a set of titles of scientific articles and chose the most interesting ones via a web form. The test subjects were shown 80 lists, with six article titles each. The subjects participating in the experiment were researchers in vision research, artificial intelligence, and machine learning.

---

[2]Note that in text modeling this corresponds to sampling a "topic" $Z$ for each word or token. In such a framework each document has a multinomial distribution $\text{Mult}(\boldsymbol{\theta})$ over topics.

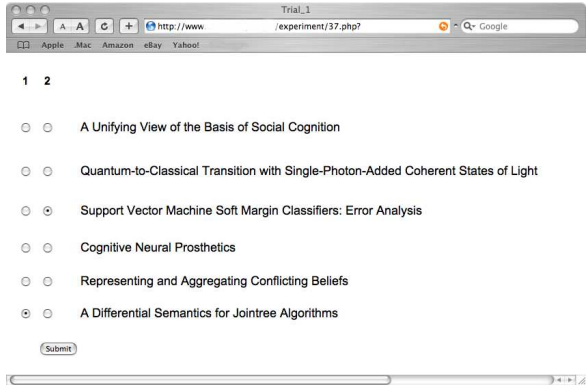

Figure 3: Web Form of the Scientific Articles Data. The test subjects were shown 80 lists, each with six article titles. On each page the task of the subjects was to choose the two most interesting titles. They were asked to give their feedback explicitly via web forms, like the one shown in the figure.

The browsed lists consisted of titles of scientific articles published during autumn 2004 in major journals in the fields of vision research, artificial intelligence, machine learning, and general science. On each page there was a randomly generated list of titles always containing titles from each discipline. On each page the subjects were to choose the two most interesting titles according to their own preferences (see Figure 3). Altogether, the data consisted of 25 users' opinions on 480 titles of scientific articles ("documents").

## 4 EXPERIMENTS

### 4.1 "NEW" DOCUMENTS AS TEST SET

We constructed the experiments to correspond to prediction of relevances for "new" documents having only few relevance judgments available. For the test set, we randomly selected $N_t$ documents to represent the "new" documents ($N_t = 50$ in the parliament data and $N_t = 48$ in the scientific articles data). We took care that these documents were represented with very few ratings in the training data, which would be the case for new documents in information retrieval. Of the ratings for these documents, randomly selected 3 ratings per document were taken to the training set and the rest of the ratings were left to the test set. For the rest of the documents, all the ratings were included in the training set. These documents represented the "older" documents that many users have already seen and revealed their opinion on. They correspond to older documents in an information retrieval situation. This way all the tested documents are "new" in the sense that we only know 3 users' opinions of them. However, we are able to use "older" documents (for which users' opinions are already known) for training the user groups and document clusters.

For the validation of parameters we used the training set to construct a validation set and a preliminary training set in a similar manner. "New" documents included in the test set were not used in the validation. From the rest of the documents we again randomly selected $N_v$ documents to be the "new" documents of the validation set ($N_v = 50$ in parliament data and $N_v = 42$ in the scientific articles data). Of the ratings for these documents, randomly selected 3 ratings per document were taken to the preliminary training set and the rest of the ratings were left to the validation set.

Our two-way grouping model and Gibbs URP were trained with the preliminary training set for a range of cluster numbers. The trained models were tested with the validation set, and the lowest perplexity (see the next section) was used as performance criterion for choosing the cluster numbers. For the final results the models were trained with all the training data with the validated cluster numbers and tested with the test data set.

### 4.2 MEASURES OF PERFORMANCE

For all the models, except the naive model, we used log-likelihood of the test data set as a measure of performance, written in the form of perplexity[3]

$$\text{perplexity} = e^{-\frac{\mathcal{L}}{N}} \text{ , where } \mathcal{L} = \sum_{i=1}^{N} \log P(r_i \mid u_i, d_i, \mathcal{D}) \text{ .}$$

Here $\mathcal{D}$ denotes the training set data, and $N$ is the size of the test set. Gibbs sampling gives an estimate for the table of relevance probabilities over all $(u, d)$ pairs, $P(r \mid u, d, \mathcal{D})$, from which the likelihood of each test pair $(u_i, d_i)$ can be estimated as $P(r_i \mid u_i, d_i, \mathcal{D})$.

We further computed the accuracy, that is, the fraction of the triplets in the test data set for which the prediction was correct. For the naive model the prediction accuracy was the only performance measure used, since, unlike the other models, it does not produce probability for the relevance. For the other models we took the predicted relevance to correspond to

$$\arg \max_{r \in \{0,1\}} P(r \mid u, d, \mathcal{D}) \text{ .}$$

---

[3] The best possible performance yields perplexity = 1 and random guessing (coin flipping) yields perplexity = 2. If perplexity is greater than 2 the model is doing worse than random guessing. Theoretically, it can grow without a limit if the model predicts zero probability for some item in the test data set. However, we actually clipped the probabilities to the range $[e^{-10}, 1]$ implying maximum perplexity of $e^{10} \approx 22,000$

## 4.3 CONVERGENCE

Convergence of the Markov chain Monte Carlo simulations was measured as explained below. We sampled three MCMC chains in parallel and monitored the convergence of predictions. First, each of the chains were run for 100 iterations of burn-in, with tempering like in (Koivisto, 2004). After that, the burn-in period was continued for another 50 iterations without the tempering, to burn in the actual posterior distribution. At the end of the burn-in period, the squared Hellinger distance[4] $H^2$ was used as a convergence check: it was required to achieve the limit of $10^{-3}$. After the burn-in each chain was run for another 400 iterations, and finally those 1200 samples were averaged to estimate expectations of $P(r \mid u, d)$.

## 5 RESULTS

The results of comparing the perplexities of the models on test data set are shown in Tables 2 and 3. Statistical significance was tested with the Wilcoxon signed rank test.

Our model performed better than URP, the reason probably being that in URP the number of "bins" is large compared to the number of training data samples. URP does not generalize over documents, so for each test document it has $K_U = 2$ bins and 3 data samples. For each test document there are only 3 training samples to estimate the parameters of the $K_U$ bins, so when $K_U$ grows the performance of URP gets inevitably worse. In our model, there are $K_U$ bins per document cluster, not per document, which makes it more robust to variation caused by the small number of training data samples.

### 5.1 PARLIAMENT DATA

The choices from which the cluster numbers were selected using a validation set for the parliament data were $K_U \in \{1, 2, 5, 10, 20, 50\}$ for the user groups and $K_D \in \{1, 2, 3, 4, 5, 10\}$ for the document clusters. The best numbers were $K_U = 2$ for URP, and $K_U = 2$ and $K_D = 2$ for the two-way grouping model. These values were used in the experiments with the test set. The results of the experiments are summarized in Table 2.

---

[4]We evaluated the squared Hellinger distance $H^2(p, q) = 1 - \sum_x \sqrt{p_x}\sqrt{q_x}$ pairwise, between the conditional distributions $P(r \mid u, d)$ produced by the 3 chains. The average of the Hellinger distances between the chains is an upper bound for the expectation of the Hellinger distance between the true distribution and the distribution obtained by the MCMC approximation.

Table 2: Parliament Data. Comparison between the models by perplexity and prediction accuracy over the test set of 50 documents. All the values differ from each other statistically significantly with P-value $\leq 0.01$. Small perplexity and large accuracy is better.

| Method | Perplexity | Accuracy % |
|---|---|---|
| Two-Way Model | 1.37 | 92.6 |
| Gibbs URP | 1.47 | 88.8 |
| Doc.Freq.Model | 4.87 | 67.8 |
| Naive Model | – | 50.4 |

### 5.2 SCIENTIFIC ARTICLES DATA

The choices from which the cluster numbers were selected using a validation set for the scientific articles data were $K_U \in \{1, 2, 3, 4, 5, 10, 15, 20\}$ for the user groups and $K_D \in \{1, 2, 3, 4, 5, 10, 15, 50\}$ for the document clusters. The parameter validation yielded cluster number $K_U = 2$ for URP and cluster numbers $K_U = 5$ and $K_D = 3$ for the two-way grouping model. Generalization over documents enables the two-way model to produce a more fine-grained description of the users with five user clusters. These values were used in the experiments with the final test set. The results of the experiments are summarized in Table 3.

Table 3: Scientific Articles Data. Comparison between the models by perplexity and prediction accuracy over the test set of 48 documents. The perplexity of the two-way grouping model differs from all the other models statistically significantly with P-value $\leq 0.01$. Small perplexity and large accuracy is better.

| Method | Perplexity | Accuracy % |
|---|---|---|
| Two-Way Model | 1.71 | 75.3 |
| Gibbs URP | 1.74 | 73.2 |
| Doc.Freq.Model | 3.76 | 69.2 |
| Naive Model | – | 67.1 |

#### 5.2.1 Clusters

The latent groups are unidentifiable, so a group number $k$ could refer to different groups during different Gibbs iterations. It is possible, however, to define clusters based on same users sharing the latent user group frequently, and analogously for the documents. (Technically, we compute the expectation of the Hellinger distance between the users' distributions over user groups.) As a result of this analysis we found the groupings for users and documents that we expected in the scientific articles data set. The clusters roughly corresponded to the users' research fields and disciplines of the journals, respectively.

# 6 DISCUSSION

We have built a latent grouping framework that can be used to predict the user-specific relevance of an unseen document, and vice versa. The model assumes that both users and documents have a latent group structure. The predictions were computed with Gibbs sampling. We compared the model against a state-of-the-art method, the User Rating Profile model (URP), also estimated with Gibbs sampling.

The task was to predict users' subjective relevances for new documents, with only very few existing ratings—still being able to utilize information about relevances of earlier seen documents for a mass of users. It resembles a collaborative filtering situation where relevance of a new document is predicted. Our method gives more accurate relevance predictions than URP in this task. In the task the available information about the attitudes of users towards the new documents is very limited, and generalizing over similar documents is beneficial. The document group structure of our model enables such generalization.

The aim of this study was to show that this kind of model can give more accurate results than models that generalize only over users. We have not yet optimized the current implementation for the requirements of concrete applications at this stage. A remaining question of future work is how to implement the model as efficiently as possible. The complexity of computing one iteration of Gibbs sampling with our model is $\mathcal{O}(N(K_U^2+K_D^2)+N_U K_U+N_D K_D+K_U K_D)$, whereas for URP the complexity is $\mathcal{O}(NK_U^2+N_U K_U+N_D K_U)$ per iteration. The number of observed $(u,d,r)$ triplets is denoted by $N$ and it dominates the complexities of both the models. However, the problem could be solved approximately more efficiently, for example as follows: One could keep a set of randomly selected samples and use it to represent all other users, while sampling only the parameters of the new incoming user.

The model could be easily extended to handle multiple-valued ratings, by replacing the binary output with a multinomial. Binary responses, however, have the clear advantage that the ratings have a natural ordering: From the posterior we obtain simple probabilities varying continuously between the extremes, 0 and 1. In comparison, the multinomial distribution does not take the ordering of the ratings into account. A multi-valued response variable would therefore probably require extra structure to take the natural ordering of the ratings into account.

The model is constructed to make its expansion by adding other sources of relevance feedback easy. Ongoing work includes combining users' explicit feedback with implicit relevance estimates gained from users' eye movements.

**Acknowledgments**

This work was supported by the Academy of Finland, decision no. 79017, and by the IST Programme of the European Community, under the PASCAL Network of Excellence, IST-2002-506778. This publication only reflects the authors' views. The authors would like to thank people in the PRIMA project for useful discussions, and acknowledge that access rights to the data sets and other materials produced in the PRIMA project are restricted due to other commitments.

**Appendix**

In this appendix we give the Gibbs sampling formulas for the posterior distributions for each variable relating to the user clusters. The formulas are analogous for document clusters. In our notation $n$ denotes an index for the observed triplets ($n \in \{1, 2, \ldots, N\}$) and $\psi$ denotes all the parameters of the model.

Sampling formula for user group $u^*$:

$$P(u_n^* \mid u_n, r_n, \psi) \propto \quad (2)$$

$$\frac{\beta_U(u_n^*)_{u_n} \, \theta_R(u_n^*, d_n^*)_{r_n} \, \theta_U(u_n^*)}{\sum_{u^*} \beta_U(u^*)_{u_n} \, \theta_R(u^*, d_n^*)_{r_n} \, \theta_U(u^*)}$$

Sampling formula for the user groups vs. users matrix $\boldsymbol{\beta}_U(u^*)$:
$$P(\boldsymbol{\beta}_U(u^*) \mid \{u_n\}, \psi) \propto \quad (3)$$

$$\mathrm{Dir}(nu^*u1 + \alpha_U(u^*)_1, \ldots, nu^*uN_U + \alpha_U(u^*)_{N_U})$$

where $nu^*uq = \#\{\text{Samples with } u_n^* = u^* \wedge u_n = q\}$.

Sampling formula for the user group probability parameters $\boldsymbol{\theta}_U$:

$$P(\boldsymbol{\theta}_U \mid \{u_n^*\}, \psi) \propto \quad (4)$$

$$\mathrm{Dir}(nu^*1 + \alpha_{u^*}(1), \ldots, nu^*K_U + \alpha_{u^*}(K_U))$$

where $nu^*k = \#\{\text{Samples with } u_n^* = k\}$.

Sampling formula for document cluster $d^*$, the document cluster vs. documents matrix $\boldsymbol{\beta}_D(d^*)$, and document cluster probability parameters $\boldsymbol{\theta}_D$ can be derived analogously ($u \leftrightarrow d$).

Sampling formula for the group-wise probability of relevance $\theta_R(u^*, d^*)$:

$$P(\theta_R(u^*, d^*) \mid \{r_n\}, \psi) \propto \quad (5)$$

$$\mathrm{Dir}(\alpha_R(0) + nu^*d^*0, \ \alpha_R(1) + nu^*d^*1)$$

where

$$nu^*d^*r = \#\{\text{Samples with } u_n^* = u^* \wedge d_n^* = d^* \wedge r_n = r\}.$$